\documentclass[12pt]{article}
\usepackage{setspace, graphicx, color,epsfig,graphics,amssymb,amsmath,subfigure,verbatim,lscape,soul,tabu}
\usepackage[colorlinks]{hyperref}

    \def \e {{\varepsilon}}

    \def \o {{\theta}}
    \def \Oo {{\Theta}}

    \def \D {{\Delta}}
    \def \d {{\delta}}

    \def \s {{\sigma}}

\newcommand{\beq}{\begin{equation}}
\newcommand{\eeq}{\end{equation}}
\newcommand{\beqr}{\begin{eqnarray}}
\newcommand{\eeqr}{\end{eqnarray}}
\newcommand{\beqrn}{\begin{eqnarray*}}
\newcommand{\eeqrn}{\end{eqnarray*}}
\newcommand{\beqn}{\begin{equation*}}
\newcommand{\eeqn}{\end{equation*}}
\newcommand{\bei}{\begin{itemize}}
\newcommand{\beii}{\begin{itemize} \item}
\newcommand{\eei}{\end{itemize}}
\newcommand{\bmei}{\begin{itemize} \compactlist}
\newcommand{\emei}{\end{itemize}}
\newcommand{\ben}{\begin{enumerate}}
\newcommand{\een}{\end{enumerate}}
\newcommand{\bes}{\begin{small}}
\newcommand{\ees}{\end{small}}
\newcommand{\bec}{\begin{center}}
\newcommand{\eec}{\end{center}}

\begin{document}
\title{Dynamic signal tracking in a simple V1 spiking model}
\author{Guillaume Lajoie\footnote{University of Washington's Institute for Neuroengineering, Seattle, WA, glajoie@uw.edu. This research was supported in part by the Washington Research Foundation's Innovation Fund.} \quad and \quad 
Lai-Sang Young\footnote{Courant Institute of Mathematical Sciences, New York
University, NY 10012, lsy@cims.nyu.edu. This research was supported in part by 
the National Science Foundation, DMS-1363161}}
\date{\today}

\maketitle

\noindent {\bf \large Abstract.} This work is part of an effort to understand 
the neural basis for our visual system's ability, or failure, to accurately track
moving visual signals. We consider here a ring model of spiking neurons,
intended as a simplified computational model of a single hypercolumn of the primary visual cortex.
Signals that consist of edges with time-varying orientations localized
in space are considered. Our model is calibrated to produce spontaneous
and driven firing rates roughly consistent with experiments, and our
two main findings, for which we offer dynamical explanation on the level
of neuronal interactions, are the following: (1) We have documented
consistent transient overshoots in signal perception following signal
switches due to emergent interactions of the E- and I-populations, and (2) for continuously moving signals, we have found that
accuracy is considerably lower at reversals of orientation than when
continuing in the same direction (as when the signal is a rotating
bar). To measure
performance, we use two metrics, called {\it fidelity} and {\it reliability},
to compare signals reconstructed by the system to the ones presented,
and to assess trial-to-trial variability. 
We propose that the same population mechanisms responsible for orientation selectivity also impose constraints on dynamic signal tracking that manifest in perception failures consistent with psychophysical observations.

\newpage

\section*{Introduction}
The human visual system is remarkable, but there are limits to its ability
to accurately track visual stimuli, as confirmed in psychophysical experiments~\cite{Boff:1986p2122,Pack2008189,Burr:2011p2180}.
This is not necessarily a liability: the degradation of our ability to perceive 
distinct visual frames beyond a certain ``refresh rate" is what produces 
the illusion of continuous changes when presented with fast changing static  
scenes, as is done in cinema, television and computer monitors~\cite{Efron:1973p1883,Thorpe:1996p1852}. Our perception of visual signals captured by the retina is the result
of very complex processing by the brain that begins in the primary visual cortex (V1) and 
involves a number of higher visual cortical areas~\cite{Hubel1995Eye}. 
The signal, which can be 
thought of as encoded in spike trains of neurons, is passed from region to
region, via pathways that both feed forward and feed back, transformed at 
each stage of processing by interactions among local neuronal populations. 
Convergence and divergence of projections between regions and 
the dynamics of local interactions may offer important clues to why 
we see what we see. 

This paper contains a numerical study of local population activity in a group 
of V1 neurons
in response to time-varying dynamic signals. In an attempt to strike a balance 
between biological realism and simplicity, we use, following~\cite{BenYishai:1995p1669,BenYishai:1997p119}, a network
with a ring structure to model one hypercolumn of V1. That is to say, to focus
on the orientation selectivity of the neurons~\cite{Hubel:1959p1933,Hubel:1962p1940}, we arrange them  around a circle, 
with the angular position of the neuron corresponding to its most preferred 
orientation. Both excitatory and inhibitory neurons are represented; this is 
important, for the dynamics are driven largely by the competition 
between these two subpopulations. Realistic 
biophysical information such as network connectivity and timescales of interaction 
are incorporated whenever possible. 
As all of the neurons in this network have essentially a common visual field, 
our signals are necessarily presented at that same spatial
location. Thus unlike most other studies with moving stimuli (see e.g.~\cite{Shadlen:1998p481,Movshon:1985p2070,Pack2008189}), our ``moving stimuli"
do not move in space; they consist of single gratings, the orientation of which changes in time.

Our goal is to study the model's response to such stimuli, in terms of how well
it tracks the movements, or changes in orientation, of the signal. To quantify
system performance, we will introduce metrics to describe the {\it fidelity} of the system,  
i.e., the extent to which its response reflects the true signal, and {\it reliability},
referring to its trial-to-trial variability. We will study system performance as
a function of signal attributes, including its strength, the frequency of orientation
switches (or frame rates), the sizes of the jumps in angles
and the regularity of the signal. 
Importantly, we will attempt to provide mechanistic explanation for system
performance, that is, to connect a system's response directly to the dynamical interactions 
between its excitatory and inhibitory subpopulations. We report wave-like activity patterns in response to changing stimuli, as was originally found in a comparable (rate) model~\cite{BenYishai:1997p119} studying purely rotational stimuli.

Two novel findings are that
(1) lateral excitation acting in concert with indirect suppression contributes to constrain
the speed at which a network can track a moving signal, and (2) in addition to time
lags, overshoots occur almost invariably at signal switches (i.e. when there is a sudden change in signal orientation), and they are exacerbated
under certain conditions predictable from the underlying dynamics.

\medskip
The paper is organized as follows: In Section~\ref{main_model}, we describe our model and present details of its calibration.
In Section~\ref{signal_response}, we discuss signal-reconstruction procedures and metrics to assess their precision. Section~\ref{overshoot} is dedicated to the study of signals that contain single orientation switches, where we first describe key network mechanisms in response the changing signals. Finally, in Section~\ref{cts_signals}, we demonstrate how our network responds to continuously changing stimuli, both regularly rotating and randomly switching. We close with a discussion about the implications
of this work and future research directions that it suggests.

\section{The model}
\label{main_model}

Our model is designed to reproduce key dynamical features of neurons in a single
hypercolumn of the primary visual cortex (V1) of primates, in response to visual stimuli. 
It is not meant to be a biophysically realistic model, but is rather an attempt to capture 
only orientation-specific responses of cortical neurons. Each neuron in our model is parametrized 
by an angle $\o$ between $0$ and $\pi$ representing its most preferred orientation, 
as in the {\it ring model} introduced by~\cite{BenYishai:1995p1669} and~\cite{BenYishai:1997p119} and studied in many
other papers e.g.\cite{Bressloff:2000p1671} and more recently~\cite{Rubin:2015p1673}.

\subsection{Model description}
\label{model}
We use spiking point neurons with exponential integrate-and-fire dynamics (EIF)~\cite{FourcaudTrocme:2003p552}. There are $N$ cells, $75\%$ of which are excitatory (E) and the rest are inhibitory (I). We will discuss below separately the
dynamics of individual neurons and network architecture. 

\bigskip \noindent
{\bf Dynamics of individual neurons}
\smallskip

The state of neuron $i$ is described
by its membrane potential or voltage $V^i$, the dynamics of which are
governed by the following equation:
\beq
\label{sys}
C\frac{dV^i}{dt}=-g_L(V^i-E_L)+g_L\D T \exp(\frac{V^i-V_T}{\D T})+I^i_{in}\ .
\eeq
The units are in mV and ms, and the constants are as usual: $C$ is membrane
capacitance, $g_L$ is a leak conductance, $E_L$ is the leak reversal potential, 
$V_T$ is a spike generating threshold and $\D T$ is the spike slope factor. Once the voltage escapes and starts to blow up, we manually stop it at $V=-30$ mV, record a spike time and reset $V^i$ to $V_r< V_T$,
where the voltage is held for a refractory period $T_{ref}$. The values 
of of the constants used in Eq. (\ref{sys}) are taken from  \cite{FourcaudTrocme:2003p552}.

The term $I^i_{in}$ describes the total input to cell $i$. It includes synaptic inputs
$s_i$ from other cells within the network, a background drive term $I^i_{bkgd}$ representing
inputs from within the nervous system, 
and an external input in the form of a signal:
\beq \label{I}
I^i_{in}=s_i+w^i_{ext}(a_{sig}I^i_{sig}+I^i_{bkgd})\ .
\eeq 
The synaptic input current is given by
\beq
s_i(t)=\sum_{j, \, t^{spike}_j} w_{ij}S_{XY}(t-t^{spike}_j-T^{XY}_d).
\label{synap}
\eeq
with
\beqn
S_{XY}(t)=
\begin{cases}
\frac{1}{\tau_{XY}}\exp(-t/\tau_{XY}) & ;  t \ge 0\\
0 &; t<0\\
\end{cases}
\eeqn
where $w_{ij}$ is the coupling weight from cell $j$ to cell $i$, and $\{t^{spike}_j\}$ are
all the spike times of cell $j$. 
We assume $w_{ij}$ depends only on the neuron types of $i$ and $j$.
More precisely, there are four values, $w_{XY}, \ X,Y \in \{E, I\}$ (Excitatory, Inhibitory).
These numbers are signed: $w_{EE}, w_{IE}>0$, and $w_{EI}w_{II}<0$. 
We set $w_{ij}=0$ if neuron $j$ does not synapse on neuron $i$, 
and if it does, then $w_{ij}=w_{XY}$ if neuron $i$ is of type $X$ and neuron $j$ is
of type $Y$. Similarly, we use $\tau_{XY}$ to denote the synaptic time constant from neuron class $Y$ to 
neuron class $X$, and $T^{XY}_d$ the (short) delay in transmission or rise time. 
Three types of neurotransmitters are considered: fast excitatory synapses 
(corresponding to AMPA) and fast inhibitory synapses (corresponding to GABA-A)
characterized by short time constants $\tau_{XY}$ of a few ms, and 
NMDA-based excitatory synapses characterized by much longer time constants 
$\tau_{nmda} \sim$100 ms. Each excitatory synaptic weight $w_{XE}$ acts on both fast (AMPA) and slow (NMDA) synaptic currents which are scaled according to the fractions $\rho^{nmda}_X\in[0,1]$. For example, for an E-to-I synapse, the coupling weight of the slow synaptic current is $\rho^{nmda}_I w_{IE}$ whereas for the fast current, the weight is $(1-\rho^{nmda}_I) w_{IE}$.

Returning to Eq (\ref{I}), the term $I^i_{bkgd}$ contains both a constant mean 
$\mu$ and a fluctuating term  taken to be white noise with variance $\e^2$,
the latter being independent across cells. This represents input from within 
the nervous system, both synaptic and modulatory, that we have not modeled.
A discussion of the signal term is postponed to the next subsection.
Finally, the pre-factor $w^i_{ext}$, drawn randomly from $[0.9,1]$ for each neuron, is 
intended to introduce heterogeneity among neurons. 

\bigskip \noindent
{\bf Network architecture} 

\begin{figure*}
\includegraphics{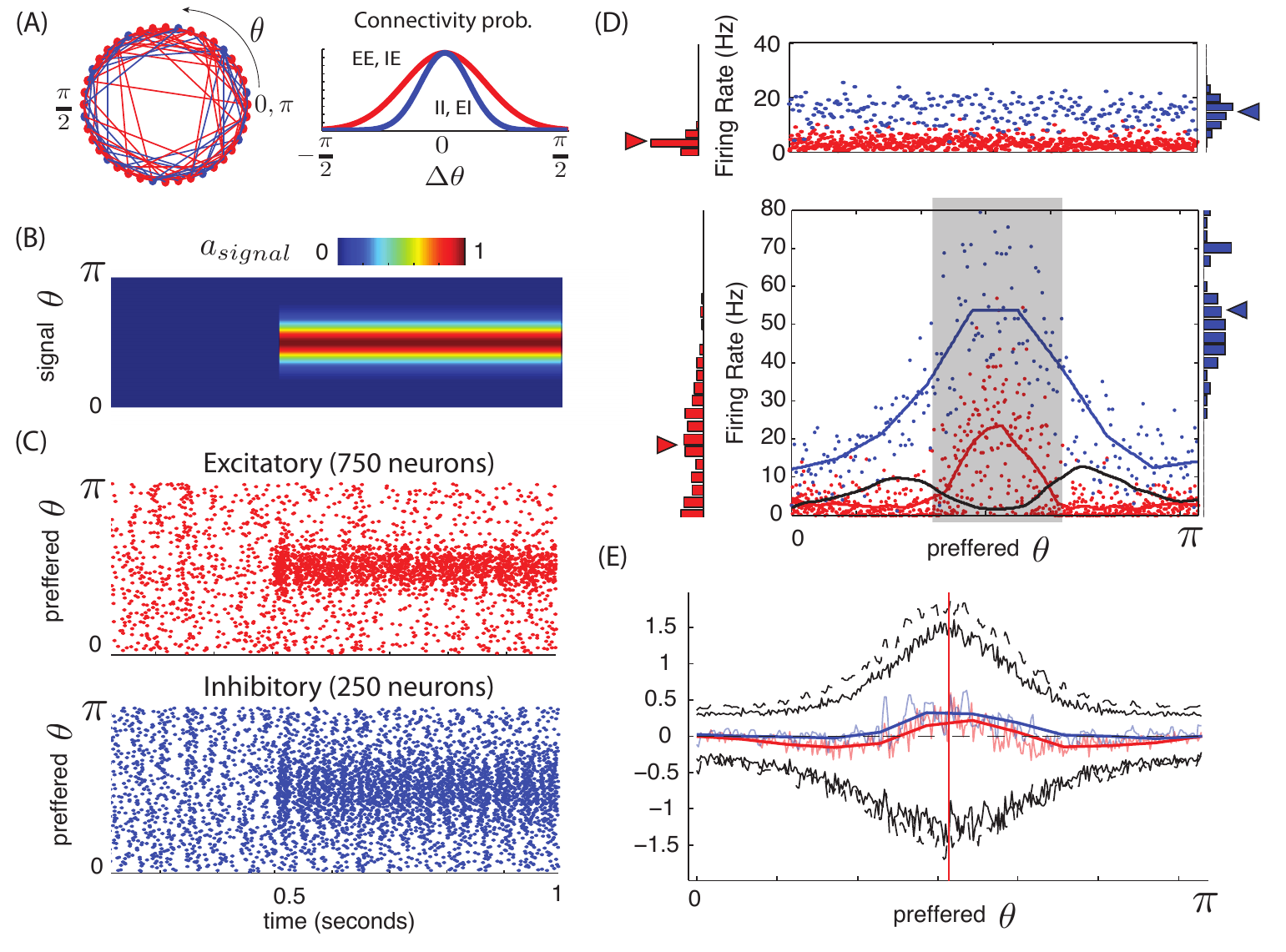}
\caption{ {\bf (A)} Network connectivity. Left: example network of $N=40$ neurons arranged in a circle with position denoting orientation selectivity ($\o\in[0,\pi]$). Right: connectivity probability distributions with respect to pre and post-synaptic type and difference of orientation preference $\D \o$ (e.g. EI means from I to E).  
{\bf (B)} Example of network signal: no preferred orientation until 0.5 seconds followed by a strong signal ($a_{sig}=1$) centred at $\o=\pi/2$ for the following half second. Colours show signal strength. 
{\bf (C)} Raster-plots showing network spiking activity separated in E/I populations in response to signal from (B). Dots indicate spike times. Top: Excitatory spikes (red). Bottom: Inhibitory spikes (blue).
{\bf(D)} Mean firing rates of each neuron in the network computed over 10 second simulations. Side histograms show firing rate distributions of E (left) and I (right) populations respectively. Arrows indicate mean. Top: in absence of stimulus (spontaneous). Bottom: in the presence of stimulus as in second part of (B). Grey box shows neurons with preferred orientation $\o$ within $\o_{sig} \pm \s_{sig}$ used to compute side histograms. Solid curves show firing rates averaged over a 20 neuron sliding window for E (red), I (blue) and the relative difference between the two (i.e. (I-E)/E, black).
{\bf(E)} Mean synaptic currents $s_i$ of each neuron in the presence of a strong stimulus as in (B), averaged over 10 seconds. In black, E (positive) and I (negative) currents received by E and I neurons, respectively solid and dashed lines. Thin red and blue lines show the difference between E and I synaptic currents, for E and I neurons respectively. Thick red and blue lines show differences averaged over sliding 10 neuron window.}
\label{fig:intro}
\end{figure*}

\smallskip
We consider a network of $N$ neurons, $N$ being on the order of 1000.
To focus on their orientation preferences, we have elected to use a network
with a {\it ring structure}, that is to say, all excitatory and inhibitory neurons 
are placed uniformly in a circle, each neuron being identified with an angle 
$\theta$ to be thought of as its most preferred orientation 
(see Figure~\ref{fig:intro} (A)). The probability that a neuron at angle $\theta$ is
connected to one at $\theta_0$ is $p_{XY}G(\theta-\theta_0)$ where
$G$ is a Gaussian with SD $\sigma_{XY}$, $p_{XY}$ representing an orientation-dependent connection probability,
and the pre- and postsynaptic neurons are of types $Y$ and $X$ respectively. 

In agreement with known experimental measurements (\cite{Beaulieu:1992p1763,Fitzpatrick:1985p1776}), the numbers $\sigma_{EE}=\sigma_{IE}$ representing the extent of axonal trees
of excitatory neurons, are larger than  $\sigma_{EI}=\sigma_{II}$, the
corresponding reach for inhibitory neurons (see Figure~\ref{fig:intro} (A)). Also following experiments, $p_{EE}$ is taken to be $0.15$, significantly smaller than all other $p_{XY}$, which we have taken to be $0.5$~\cite{Oswald:2009p2183,Holmgren:2003p2182}.

We stress that our networks are randomly drawn according to the connection
probabilities above. Once a graph corresponding to a realization of the network is
chosen, we fix it for the duration of
the study. We then verify that the results we obtain are not dependent of connectivity realization. 
The simulations we show in this paper are for $N=1000$, and we have checked
that qualitative features of our results persist for networks of size $N$ ranging
from $\sim 500$ to a few thousands.

\subsection{Model calibration} 
\label{calibration}
Values of all of the parameters used are given in the Appendix. They are determined as follows: We use  biophysical guidance when we can. As for the rest, which includes in particular
the coupling weights $w_{XY}$ and values associated with $I^i_{bkgd}$, 
they are determined by tuning parameters to produce firing rates that are 
consistent with experimental data (e.g.~\cite{Ringach:2002p1807,Ringach:2002p1805,Ringach:2002p1803}).

Figure~\ref{fig:intro} shows various basic features of our model. The top panel
in (D) shows
spontaneous firing rates at 3-7 Hz for E and 10-20 Hz for I in a realization 
of our model network, consistent with experiments. 
Panel (B) shows the onset of a strong signal favoring a specific orientation 
(e.g. a grating), and Panel (C) contains rasters showing the elicited response 
in our model neurons. Observe the gamma rhythms, which are especially 
apparent in the I-population. These rhythms are an entirely emergent phenomenon, and the 40-50 Hz frequency mimics observations in cortex~\cite{Henrie:2005p1830,Gray:1989p1811}.
The bottom panel in (D) shows firing rates under the strong center drive in (B), 
confirming the significantly higher firing rates
for neurons whose preferred orientations agree with that of the signal.
Finally, the top and bottom black curves in Panel (E) show the total synaptic
E- and I-currents 
received by individual neurons as functions of $\theta$. One can see, for example,
that this is by and large a balanced network, a hallmark of cortical activity (see e.g. discussion in~\cite{vanVreeswijk:1996p476}).

\section{Responses to signals}
\label{signal_response}
This section discusses some steps used to filter the response to produce a reconstructed signal, and introduces some metrics aimed at evaluating the model's ability to track a signal properly.
Throughout, we use numerical simulations of model~\eqref{sys} to sample statistics about its response properties to orientation signals. More precisely, we draw
randomly a network realization, and study its response properties to signals of
various types. We then verify that our conclusions are not dependent on specific realizations. 
More details about numerical methods can be found in the Appendix.

\subsection{Signals and their reconstruction}
\label{reconstruct} 
As all of our neurons are assumed to share a localized receptive field, the stimuli
we use mimics one at a fixed location in visual space. We will refer to a signal 
$\theta_{sig}(t)  \equiv \theta_0 \in [0, \pi)$ as a {\it constant signal}; think of it
as a drifting grating with a fixed orientation. We are primarily interested, however,
in dynamic signals, as in sequences of edges with orientations which vary with time. The functions $\theta_{sig}(t)$ we consider are
mostly piecewise constant in time.

At any one instant, the impact of the signal on the network is 
represented by a Gaussian-shaped function $\Phi(\o)$
centered at $\theta_{sig}$ with $\sigma_{sig} = \pi/10$, adjusted by a multiplicative constant so $\Phi(\theta_{sig})=1$. 
The value of $I^i_{sig}$ in Eq. (\ref{I}) is the value of $\Phi$ at  the angle 
occupied by neuron $i$, scaled by the signal strength pre-factor $a_{sig}$,
which can be thought of roughly as {\it contrast} though the correspondence
is nonlinear. The Gaussian shape of $\Phi$ models the fact 
that neurons in an afferent layer are sensitive to a range of orientations.
From the parameter tuning discussed at the end of Section~\ref{calibration}, we have 
found that $a_{sig} \in [0,1]$ captures an adequate range, leading to realistic firing rates, with $a_{sig}=1$
corresponding to the strongest signal. Both E and I neurons are affected 
in the same way by the signal.

\bigskip \noindent
{\bf Filters and ``read-out" functions}

\smallskip
Downstream areas in the visual system integrate information from the 
spiking activity of neurons in each layer. The precise mechanisms enabling this readout falls far outside of the scope of this model. Here, we adopt a heuristic approach, and 
construct what we believe to be a reasonable filter to summarize the spiking activity of the E-population in response to 
a signal. Summarizing statistics are necessary to enable simple, informative 
metrics aimed at exploring the system's ability to track time-dependent signals 
under a range of conditions. 

Let $\mathcal E \subset \{1,2,\dots, N\}$ denote the set of all indices $i$
corresponding to E-neurons, and let $\o_i=i\pi/N$. 
Suppose upon the presentation of a stimulus (or simply in background) that 
excitatory firing is comprised of the following collection of spike trains 
$$
r_{\o_i}(t)=\sum_{t^{spike}_i} \d(t-t^{spike}_i), \qquad i \in \mathcal E, \ \ t \ge 0\ .
$$
Consider the filter or read-out function  $R(\o,t)$ defined by
\beq
\label{readout}
R(\o,t)=\sum_{\o_i=0}^\pi G(\o-\o_i)\int_{-\infty}^tH(t')r_{\o_i}(t')dt'
\eeq
where $G(\o)$ is a (periodic) wrapped-Gaussian filter in orientation space $[0,\pi]$ centered at $\o=0$ with variance $\s_G^2$, $H(t)$ is a half-gaussian causal filter in time centered at $t=0$ with variance $\s_H^2$, with gradual decay for $t'<t$ and zero for $t'>t$. Readout filtering parameters are set to $\s_H=40$ ms and $\s_G=\frac{\pi}{10}$ radians. Temporal delays in signal integration by V1 cells have been 
reported to be in a comparable range~\cite{Hawken:1996p2188} and
we verified that varying the filters' widths within reasonable ranges did not qualitatively affect our results.

While there are many techniques developed to analyze the spiking activity of neural populations (see e.g.~\cite{abbott1999neural,Gabbiani:2009p47}), our goal is not to derive an optimal estimate of encoded signals, but rather to mimic the information received by a down-stream population. The temporal filter $H$ represents the population's integration time-constant (see e.g.~\cite{Ringach:1997p1916}) while the spatial filter $G$ represents the breath of orientation selective projection.
The function $R(\o,t)$ can be seen as a summary of the state of the system 
as indicated by its very recent spiking activity. More than for rate models, 
a summary statistic of this kind is necessary for spiking neuron models, as 
actual spike trains are unwieldy to work with.
We expect that read-out functions of this type have been used in other modeling work, 
but were unable to locate suitable references.

\begin{figure*}
\includegraphics{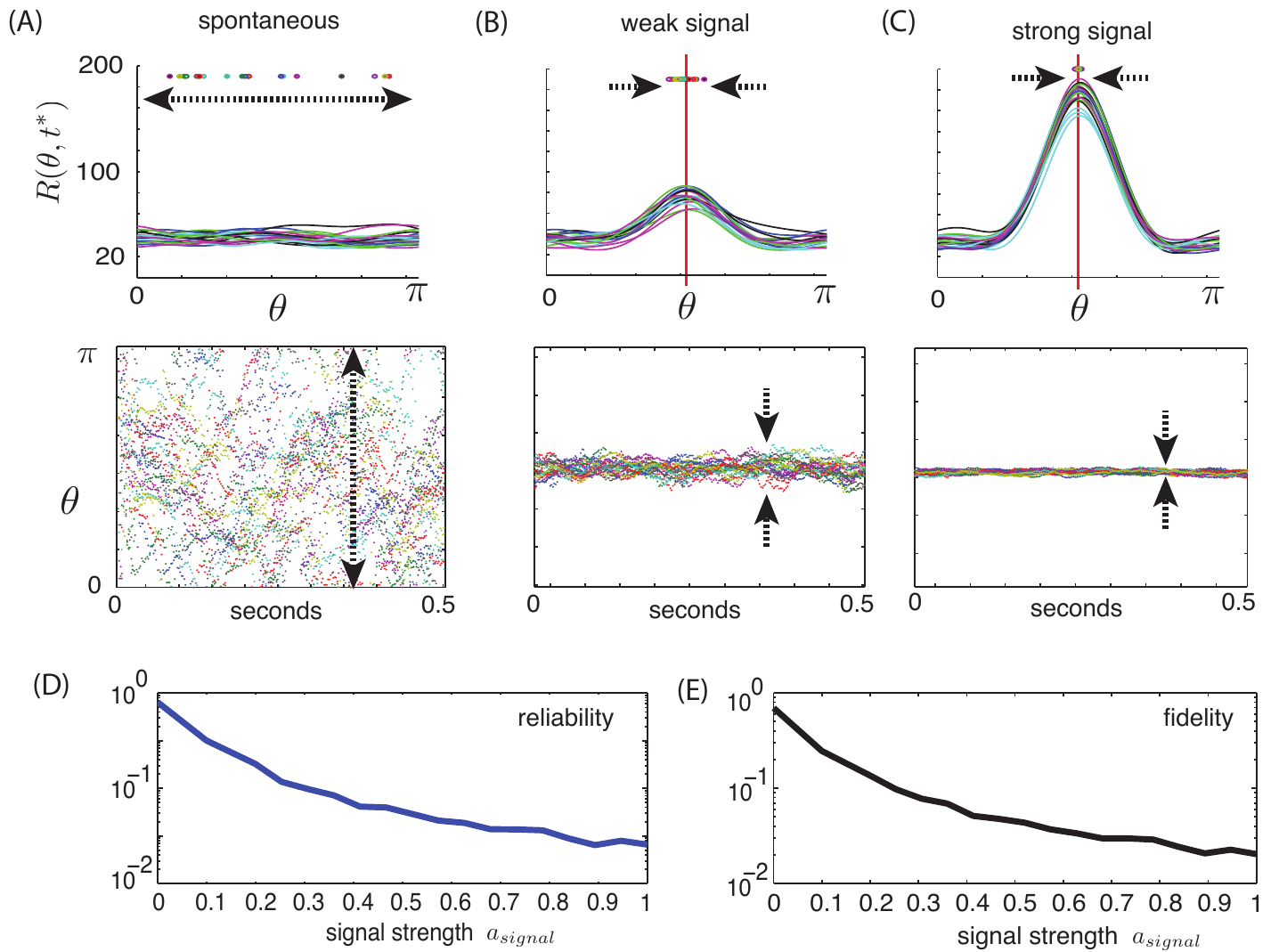}
\caption{{\bf(A, B, C)} Top row: snapshots of activity profiles on 20 distinct trials. Lines indicate trials' profiles $R(\o,t^*)$ at a time $t=t^*$, dots show corresponding reconstructed orientations $\Oo(t^*)$. Red line shows orientation signal when present, arrows show breath of cross-trial variability. Bottom row: temporal evolution of reconstructed signals $\Oo(t)$ (dots in top row) for 20 trials over a half second. (A) No signal: $a_{sig}=0$. (B) Weak signal: $a_{sig}=0.25$. (C) Strong signal: $a_{sig}=1$. 
{\bf(D, E)} Log-scale plot of the Reliability and Fidelity of the network response to 
a constant signal as a function of signal strength $a_{sig}$. Values are estimated for a one minute presentation over 50 trials for a range of signal strengths $a_{sig}$.}
\label{fig:readout}
\end{figure*}

Example snapshots of $R(\o,t)$ in response to a constant signal of three distinct 
strengths is shown in the first row of Figure~\ref{fig:readout}, the many
curves representing $R(\o,t)$-functions computed from different trials. 

\bigskip \noindent
{\bf Mean vector strengths}

\smallskip
For an even more compact statistic,
the estimator 
\beq
\label{reconstruct}
\begin{split}
\Oo(t)&=\text{arg} \left[\sum_{j \in \mathcal E}R(\o_j,t)e^{i2\o_j} \right]\ ,
\end{split}
\eeq
called the {\it mean vector strength}, represents the weighted mean orientation 
with respect to population activity~\cite{Georgopoulos:1986p820,Salinas:1994p1840}. We treat this quantity as the reconstructed signal by the network. Its simplicity as a single scalar makes it a natural tool for exploring our network's response property.

Dots on top of the first three panels of Figure~\ref{fig:readout} show the reconstructed values $\Oo(t)$ on several trials, for three different signal strengths. In (A), not surprisingly, the reconstructed orientation $\Oo(t)$ is random and appears roughly uniformly distributed along $[0,\pi]$ since there is no signal present. In (B) and (C) the reconstructed signals correctly reflect the
true signal, but still fluctuate due to ongoing network activity and to noise, 
especially in (B), where the signal is weak. 

\subsection{ Fidelity and reliability}

Measures of the quality of a system's performance in signal reconstructions are needed. We work with a fixed network once it is drawn, and simulate dynamics in response to a given signal on many {\it trials}. Distinct trials in the discussion to follow correspond to presentation 
of the same signal to the same network, while internal conditions of 
the network at the time of signal presentation may differ. As an abstraction, 
we assume that initial network conditions are randomly selected, and 
fluctuating background drive components during presentation are 
independently drawn from trial to trial.
Here, we introduce two metrics aimed at quantifying network performance across many trials. 

The first is intended to measure {\it fidelity},
by which we mean the ability of the system (model or real brain) to track
signals accurately.  The definition below depends not only on the signal but also on the read-out
function from which the estimator $\Oo(t)$ is computed, though we will suppress
this and express fidelity only as a function of the signal $\{\theta_{sig}(t)\}_{t\in [0,T]}$: 
$$
Fid[\{\theta_{sig}(t)\}_{t\in [0,T]}]= \frac{1}{T} \int_0^T \langle |\Oo(t)-\o_{sig}(t)|\rangle
\ dt $$
where the average $\langle \cdot \rangle$ is taken over distinct trials 
for the same signal. $Fid = 0$ or very close to $0$ for a signal means, 
according to this definition, that in almost every trial the mean vector strength never
deviates much from the true value of $\theta_{sig}(t)$. 
In the case of nonconstant signals a correction can be used to offset the
delay incorporated into the read-out function; this is discussed in Section~\ref{cts_signals}.
 
The second metric we consider is {\it reliability}, meaning trial-to-trial
variability in the system's response to repeated presentations of the same stimulus. 
We define it to be
$$
Rel[\{\theta_{sig}(t)\}_{t\in [0,T]}]= \frac{1}{T} \int_0^T \sqrt{Var(\Oo(t))} \ dt
$$
where the variance is taken over distinct trials with the same signal.

The concepts of fidelity and reliability are related in the following way: Fidelity
measures how faithfully $\Oo(t)$ reproduces the signal $\theta_{sig}(t)$, 
whereas reliability measures the variability in the ensemble of
trials about its own mean. A system can be reliable but have poor fidelity (meaning
it consistently
gives the same wrong results), whereas unreliability in general will 
result also in poor fidelity. 

Both fidelity and reliability are first and foremost properties of the network, 
though the same system can perform better for some types of signals than 
for others as we will discuss in Section~\ref{cts_signals}.
Panels (D) and (E) of Figure~\ref{fig:readout} show the reliability and fidelity 
of our model 
in response to a constant signal as a function of signal strength $a_{sig}$. 
For stronger signals, these numbers are quite small, confirming that our model 
performs well for signals that are constant in time. Their performance for time-varying signals
is the subject of Sections~\ref{overshoot} and~\ref{cts_signals}.

\medskip
Fidelity and reliability are well studied concepts in neuroscience, though
we know of no standardized definitions; see e.g.~\cite{Tiesinga:2008p462}
 for an overview in the context of spike time, or e.g.
 ~\cite{Faisal:2008p1563,Amarasingham:2006p1914} at the spike count level. There are also different levels of precision one can consider. 
The version we use here is defined in terms of $\Oo(t)$, a statistic that summarizes
population activity; it is less refined than, e.g., spike-time reliability for individual neurons,
which has been related to Lyapunov exponents of dynamical systems~\cite{Ritt:2003p251,Lin:2009p350,Lajoie:2013p785}. In the present context, we believe that filtered activity at the level of populations is a more relevant observable.

\section{Overshooting and dynamical explanation}
\label{overshoot}

We are interested in testing the temporal signal-response properties of our network in an effort to better understand the limitations of orientation perception in neural networks with an architecture inspired by visual cortex. In this section, we focus on signal switches,
i.e., at jump times for piecewise constant $\theta_{sig}(t)$.  In Section~\ref{switch_obs}, we report
on some findings surrounding an overshooting phenomenon that to our knowledge is new, and 
in Section~\ref{switch_dyn}, we propose an explanation in terms of underlying dynamical interactions between the E- and I-populations.

\subsection{Overshooting at signal switches}
\label{switch_obs}

\begin{figure*}
\includegraphics{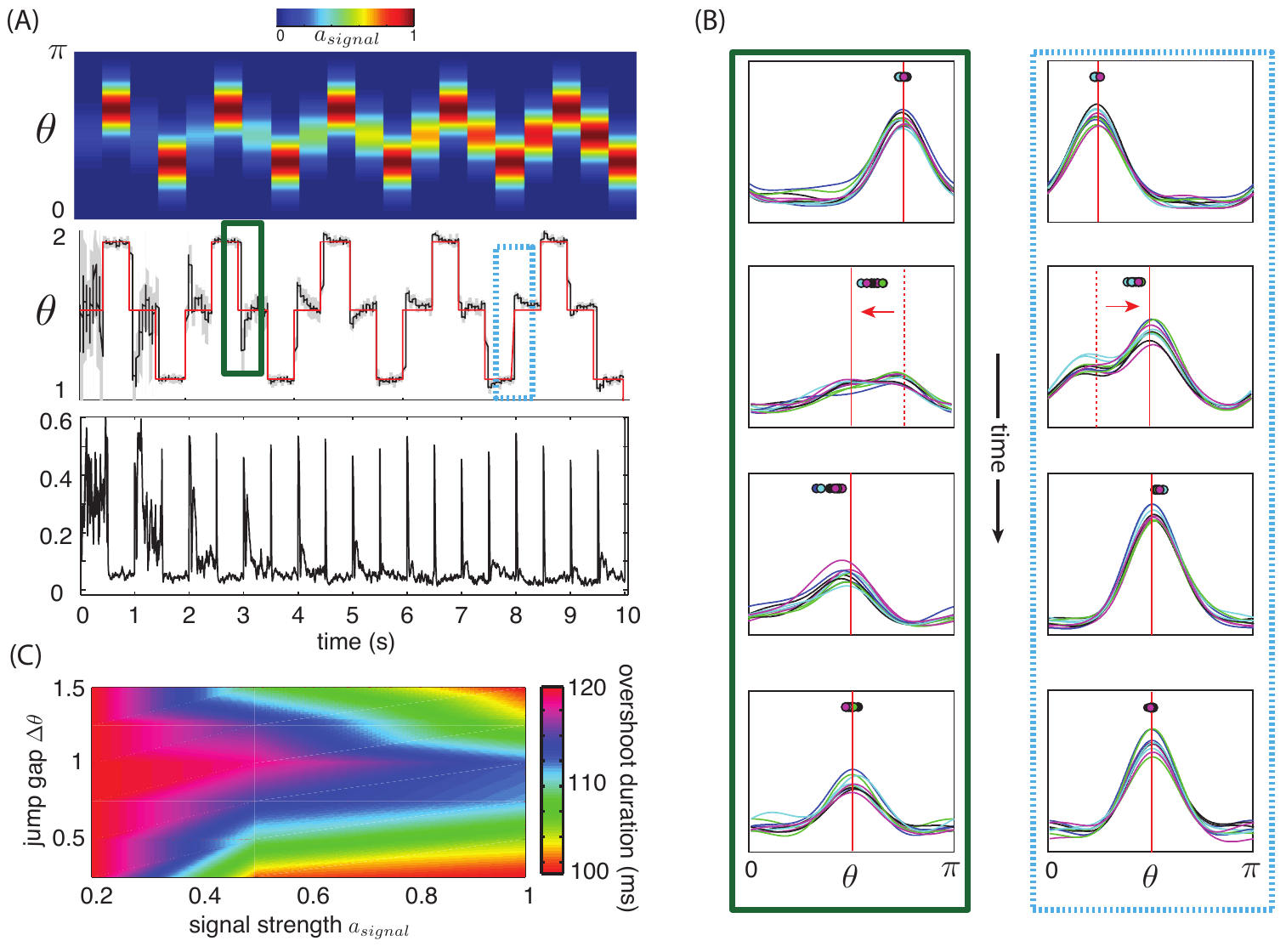}
\caption{{\bf (A)} Top: Switching signal protocol where signal keeps switching from strong side orientations ($\pi/2 \pm 0.8$) to center ($\pi/2$) with progressively increasing strength. Middle: signal reconstruction $\Oo(t)$ averaged over 50 trials in black with Reliability shown as a shaded area surrounding the curve. Red lines show underlying signal as in top panel. Bottom: Fidelity curve for $\Oo(t)$ in middle panel computed over 50 trials. {\bf (B)} Sequences of filtered activity profiles $R(\o,t)$ of 10 trials at four moments surrounding signal jumps indicated with boxes in (A). Solid coloured lines correspond to $R(\o,t)$ and dots show orientation estimates $\Oo(t)$. {\bf (C)} Color plot of reconstructed signal convergence time following a switch, from a strong to a target signal, as a function of gap size $\Delta \o$ and target signal strength. Values estimated using 50 trials undergoing 30 signals jumps for each parameter pair $(a_{sig},\Delta \o)$.}
\label{fig:gap}
\end{figure*}

Of interest here are single signal switches, by which we mean the following:
Suppose a signal switch occurs at time $t^*$. That is to say,  for $t<t^*$, 
$\theta_{sig}(t) \equiv \theta_1$ and 
$a_{sig}(t) \equiv a_1$, whereas for $t>t^*$,  $\theta_{sig}(t) \equiv \theta_2$ and 
$a_{sig}(t) \equiv a_2$. We assume $\theta_1 \ne \theta_2$, while $a_1$ may or
may not be equal to $a_2$.
Our findings, which are illustrated in Figure~\ref{fig:gap},
can be summarized as follows:

\begin{itemize}
\item Overshoots occur: If, for example, $\theta_2>\theta_1$, then our reconstructed signal $\Oo(t)$ is $ > \theta_2$ for a transient time period
 immediately following the switch.
\item  The magnitude of this overshoot depends on the relative strengths of
the signal before and after the switch: It is more prominent if $a_1 \gg a_2$, i.e. 
when one switches from a strong to a weak signal, 
less prominent but clearly present when $a_1 \approx 1 \approx a_2$,
i.e. when both signals are strong, and 
is less noticeable when $a_1 \ll a_2$. 

\item The time it takes for  $\Oo(t)$ to return to the value of $\theta_{sig}(t)$
is on the order of 50-150 ms. It is
roughly proportional to the magnitude of the overshoot and depends on the switch gap $\D \o=|\o_1-\o_2|$.
 \end{itemize}
These points are illustrated in the three panels of Figure~\ref{fig:gap}(A), where
we see clearly that fidelity suffers the most in the $\sim 100$ ms or so following
signal switches. 

Figure~\ref{fig:gap} (B) offers more detail on what happens during these switches.
It shows four snapshots of the read-out function $R(\o,t)$, which describes 
network-wide activity, for multiple trials. While there is variability, the
overall trend is robust: each of the $R(\o,t)$ profiles evolves in a wave-like fashion
with time, from having a peak at $\theta_1$ to having a peak at $\theta_2$ following
very similar routes. The solid dots at the top of each box show the values of 
$\Oo(t)$ for the different trials; they confirm that overshoots occur
for all the trials in the examples shown. The right column, for which signal
strengths before and after the switch are comparable, show a smaller overshoot
than in the left column, where $a_1$ is considerably larger than $a_2$.

Finally, Panel (C) in Figure~\ref{fig:gap} shows overshoot duration as functions 
of the final signal strength $a_{sig} = a_2$ (with $a_1\equiv 1)$ 
and switch gap $\Delta \theta = |\theta_1-\theta_2|$. 
With regard to signal strength, there are no surprises: 
the weaker the target signal, the longer the convergence time.
Of note here is that at every signal strength, the system performs better 
in terms of overshoot duration for some $\Delta \theta$ than for others, 
with the ``worst $\Delta \theta$'' occurring at $\Delta \theta \approx 0.9 \pm 0.1$ 
radian as $a_{sig} \to 1$.

To make the overshoot easily visible, $\Delta \theta=0.8$  radian is used
in Panels (A) and (B) of Figure~\ref{fig:gap}, though the phenomenon is
robust for a wide range of $\Delta \theta$. 

\subsection{Underlying dynamical mechanisms}
\label{switch_dyn}

To understand what goes on at signal switches, let us first examine the model's
response to {\it constant} signals, for this holds
the key to understanding the phenomena observed.
We begin by identifying a few relevant features, which we do not claim 
are novel or are exclusive to our model. We need to discuss first these mechanisms 
because they will be
used to explain the overshooting phenomenon described in Section~\ref{switch_obs}, as well as other
phenomena to be discussed in the next section.

Revisiting Figure~\ref{fig:intro}, we observe that
firing rates aside, the most striking difference between the responses of 
the E- and I-populations in Panel (C) to the signal in Panel (B) is the following:

\medskip \noindent
(1) {\it The interval of I-cells with elevated spiking is considerably
wider than the interval of E-cells.}

\medskip
This fact is corroborated in the second plot in Figure~\ref{fig:intro} (D), which shows a considerably
wider profile of elevated spiking for I-cells. While these are histograms of population
spiking activity (and not tuning curves of individual neurons), the two are connected
in a straightforward way: For a cell at distance $d$ from $\theta_{sig}$ to have
elevated spiking when $\theta_{sig}$ is presented means this neuron responds
to a signal at distance $d$ from its most preferred orientation (as a result of
both network and signal afferents). That is to say, its tuning curve is wide enough
to include angles at distance $d$ from where it is peaked.
 Thus the simulation results in Panel (D) of Figure~\ref{fig:intro}
are in agreement with what is generally believed to be the case for tuning curves, 
namely that E-cells are more sharply tuned, and I-cells more broadly tuned. 

We remark that in our model,
Observation (1) is an emergent phenomenon, in that our signal
affects E- and I-cells at the same $\theta$ in identical ways. 
As to why the interval of elevated spiking for I-cells is broader, we conjecture 
that this is mostly a consequence of the fact that
the differential between driven and spontaneous firing rates is larger for I-cells
than for E-cells, and the taller peak for I-cells takes a larger $\theta$-interval to 
return to baseline values; a very steep drop in I-firing rates, i.e., nearby I-neurons receiving
very similar inputs but having vastly different firing rates, seems counter-intuitive.
Another important point is that once the I-cells have an advantage,
E-cells tend to be further suppressed.
These factors are consistent with existing explanations for the sharpening of
orientation-selectivity of E-cells in cortex. See e.g. \cite{Priebe:2008p1675} for a review on the subject.

\medskip \noindent
(2)  {\it Population activity in response to a constant signal has 
a ``Mexican-hat" profile, with the largest dips in activity occurring, for the connectivity
profiles used in our model, at $\sim 0.9 \pm0.1$ radians from $\theta_{sig}$.}

\medskip
This is evident in a few of the figures shown. In Figure~\ref{fig:intro} (D), the black curve, 
which represents (I-fr -- E-fr)/E-fr ({\it fr} meaning firing rate), has two bumps on the flanks
of $\theta_{sig}$, suggesting that at these locations, the suppressive effect of the I-population
is likely to have the most significant effect on the E-population. Panel (E) in
the same figure
shows the same for synaptic currents: The red, resp. blue, curves in the 
middle represent (E-synaptic current - I-synaptic current) into E, resp. I, 
cells. Notice that the red curve again has two
valleys on the flanks of $\theta_{sig}$, where it in fact dips below zero
(this does not mean E-cells here cannot fire; 
external and background drives, which are identical for E and for I, are not included in these graphs). Similar profiles can be seen in the filtered network responses in Figure~\ref{fig:readout} (B,C).

``Mexican-hat" profiles have been observed experimentally (see e.g. \cite{Priebe:2008p1675}). In theoretical studies they are sometimes assumed for tuning curves 
of individual neurons \cite{Series:2004p575,Spiridon:2001p715,Laing:2001p1731} 
and sometimes appear as an emergent phenomenon (see e.g. \cite{Somers:1995p1672,
Kang:2003p1667}). In our model, it is entirely emergent, occurring as a result of the dynamical interaction between the E- and I-populations for reasons similar 
to those given for Observation (1).

Finally, we wish to mention one other model feature not discussed thus far: When an E-cell
spikes, the neurotransmitters are of two different kinds: AMPA (fast) and NMDA (slow);
see Section~\ref{model}. We have assumed in the model, as is generally believed to be the case in the real brain~\cite{Lisman:1998p2184,Koulakov:2002p2187,Grunze:1996p2185}, that

\medskip \noindent
(3)  {\it the NMDA-component in E-synapses is larger for postsynaptic
I-cells (0.5) than for postsynaptic E-cells (0.25).}

\bigskip \noindent
{\bf Proposed explanation for overshooting following signal switches}

\smallskip
We claim that (1)--(3) above offer at least partial explanation for the phenomena reported in
Figure~\ref{fig:gap}. Suppose the signal switches from $\theta_1$ to $\theta_2$ and,
as in the left column in Panel (B), that $\theta_2$ lies to the left of $\theta_1$. 
(All analogous statements apply to the right column in this panel, which depicts
responses to a signal that moves in the opposite direction.) 
The second box represents a snapshot after the signal has switched.
Here one can see the bump in $R(\o,t)$ attempting to follow the signal. The overshoot
in the third box is accompanied by lowered E-firing rates for $\theta \in (\theta_2,\theta_1)$ than on the far side of
$\theta_2$. We propose that the depressed firing on the interval
between $\theta_1$ and $\theta_2$ can be explained by items (1) and (2) above:
Notice that $\theta_1$ and $\theta_2$ are separated by $0.8$ radian, far enough
apart to be clearly distinguishable but not far enough so that the entire
interval between them lies in the region of elevated I-firing caused by  
$\theta_1$. After time $t^*$, this elevated I-spiking persists for some time, 
temporarily holding back E-activity even as E-cells near $\theta_2$ are now
being strong driven by the new signal. Elevated I-spiking in fact extends to the
far side of $\theta_2$, but because it decreases with distance from $\theta_1$, 
its effect on the far side is less prominent, leading to the asymmetry in $R(\o,t)$
for a certain duration after time $t^*$.
Finally, the longer-lasting NMDA-effects on I-cells, i.e. item (3), prolong 
the excitation of I-neurons in this region for another $50-100$ ms.

The explanation above is also consistent with the observation that the overshoot is
more prominent when one switches from a strong to a weak signal (as discussed
in the second bullet
at the beginning of Section~\ref{switch_obs}): The stronger I-suppression on the interval 
between $\theta_1$ and $\theta_2$ in relation to the new signal's ability to 
elevate E-spiking nearby leads naturally to a more exaggerated asymmetry 
in the $R(\o,t)$ profile. Conversely, when switching from a weak to a strong signal,
the lingering weak suppression has little effect on the new and stronger
signal's ability to arouse the E-population nearby. 

Finally, we note that the ``worst" value of $\Delta \theta$ in the sense
of overshoot duration in Panel (C) is at $\sim$ 0.9 radians, and that coincides
with the locations
of the deepest valleys observed in item (2) above. 

\medskip
We finish by noting that the dynamical analysis above goes beyond the 
overshooting phenomenon described in Section~\ref{switch_obs}. The profiles $R(\o,t)$,
a few snapshots of which are shown in Figure~\ref{fig:gap} (B), vary in a tractable
and predictable way as a function of time. They offer much insight
into how the system responds to single signal orientation changes.  The dynamics of $R(\o,t)$ are predictable because a theoretical understanding can be deduced from
 the interaction between the E- and I-populations in the model network as we have
 discussed, together with smoothing properties of the filter.


\section{Continuously varying signals}
\label{cts_signals}

Having analyzed systematically what happens at single signal switches, we 
perform in this section some stress tests on both our model's ability to track more complex signals, and our understanding
of the dynamics of $R(\o,t)$. We will
continue to use piecewise constant $\theta_{sig}(t)$, with jumps occurring at
time intervals $dt$, and study the fidelity of signals of different kinds 
as a function of $dt$ and signal strength $a_{sig}$. 
We will also investigate bounds for $a_{sig}$ and $dt$ below which the system effectively fails, and to compare these bounds to those for real visual systems.
In the electronic media, one generally speaks about {\it frame rates}, or {\it refresh rates}.
As $t$ in this paper is in ms, the number of frames per sec is given by $1000/dt$,
e.g. $25$ frames per sec corresponds to $dt = 40$ ms. 

\begin{figure*}
\includegraphics{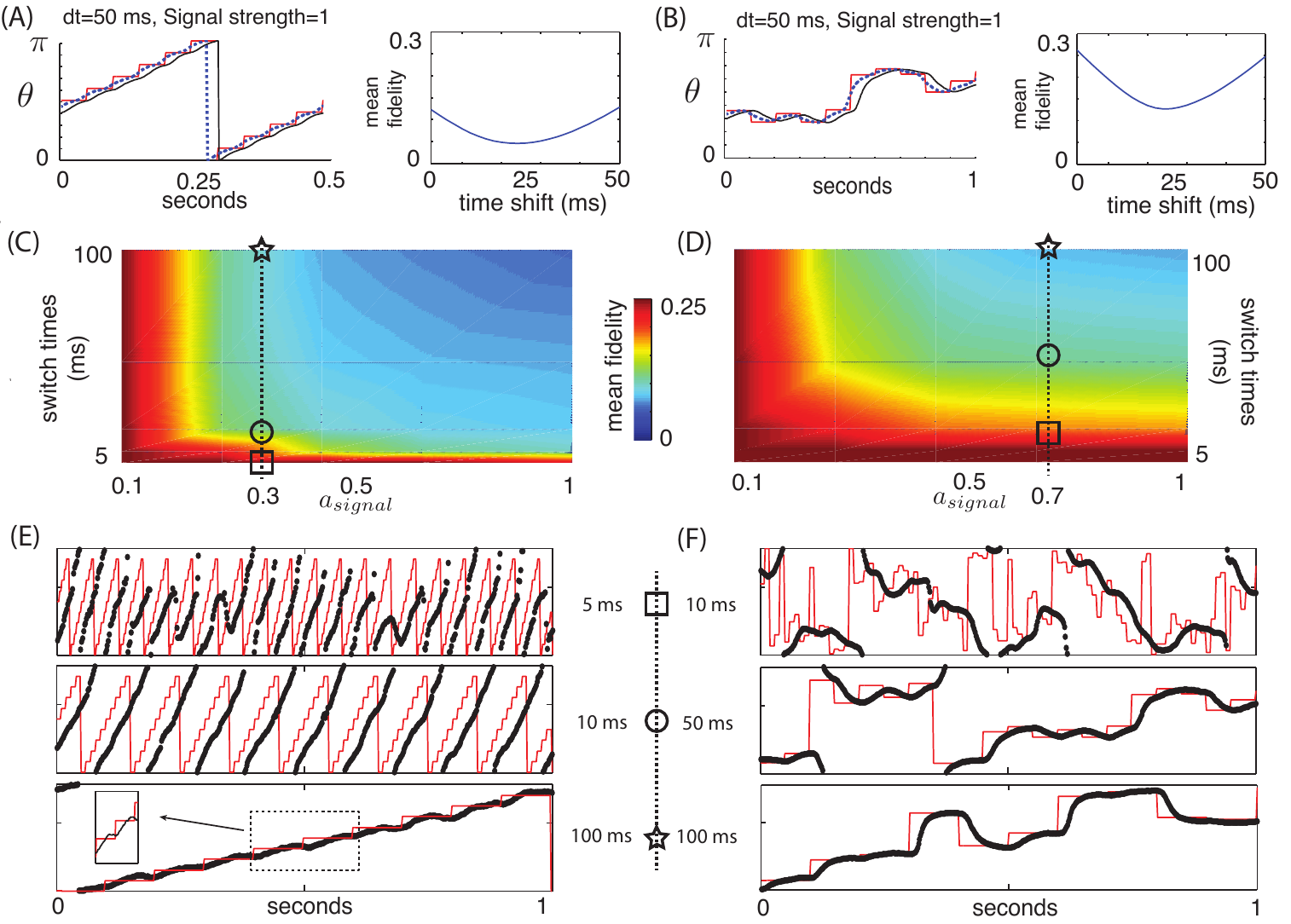}
\caption{{\bf(A, B)} Left: Mean reconstructed angles $\Oo(t)$ (computed over 30 trials of 10 seconds) in response to a strong signal ($a_{sig}=1$), composed of piecewise constant values with jumps of $\pi/10$ every 50 ms. Signal shown in red, mean response in black, shifted mean response in dotted blue minimizing the $Fid$ (error between response and signal). Right: mean fidelity $Fid$ as a function of response temporal shift. Optimal shift is at minimum. (A) shows regular rotating signal and (B), random orientation changes as described in text.
{\bf (C,B)} Colour plot of optimal mean fidelity $Fid$ (after temporal shift as in (A,B)). (C) is for regular signals and (D) for random signals (see text for details).
{\bf (E,F)} One second examples of mean response (black dots, sampled at every 2 ms) to signals (red) for different parameters indicated by symbols in (C,D). Y-axis is $[0,\pi]$ circle, meaning the top and bottom are identified. (E) is for regular signals and (D) for random signals.}
\label{fig:dyn}
\end{figure*}

Before going further, there is a technical issue we wish to take care of, namely 
there is some inherent amount of delay preventing the system from tracking
any signal perfectly if we compare $\Oo(t)$ to $\theta_{sig}(t)$ for the same 
value of $t$. This delay is due in part to the rise times of E-neurons but 
the bulk of it comes from our filter $R(\o,t)$. As we are not especially interested
in this time delay, we introduce a notion of fidelity that corrects for it,
by looking for the optimal $s$ 
with the property that the readout $\Oo(t)$ best fits the time-shifted signal 
$\theta_{sig}(t-s)$. 
For most signals, there appears to be an optimal time shift, usually between
20-30 ms, as illustrated in
the two examples in the top panels of Figure~\ref{fig:dyn}. This corrected version of fidelity,
i.e., one that has incorporated into it a time delay, will
be used from here on. 

\bigskip \noindent
{\bf Regular vs random signals}

\smallskip
We consider the following 2 types of signals: {\it regularly rotating}
and {\it random switching}, the latter intended to some degree as a model of
``natural scenes". For regularly rotating signals, $\theta_{sig}$ jumps
by a fixed amount, $\Delta \theta$, in a fixed direction at time intervals $dt$. 
For all simulations presented in Figure~\ref{fig:dyn}, $\Delta \theta= \pi/10$. For randomly switching signals,
each time the signal is refreshed, the jump can be in either direction with
equal probability, and the magnitude of the jump $\Delta \theta$ is a random variable which, for definiteness
we fix as follows: with probability $\frac12$, $\Delta \theta \in (0, \pi/10)$ with 
the uniform distribution, and with probability $\frac12$, it is drawn from a distribution
whose density is supported on $(\pi/10, \pi/2)$ and decreases linearly from
$\pi/10$ to $0$ at $\pi/2$.  
Heuristically we think of the part of the density in $(0, \pi/10)$ as due 
to small eye or head movements of
the subject, and the part on $(\pi/10, \pi/2)$ as due to genuine changes in ``scenery"
in the receptive field of the neurons in question. For simplicity,
we will refer to these two types of signals as ``regular" and ``random". 

A motivation for these choices of signals is that our analysis
in Section~\ref{switch_dyn} predicts that our model will react quite differently
to them.

\bigskip \noindent
{\bf Our findings} 

\smallskip
Numerical simulations of our model's responses to these two types of signals
for $dt \in [5,125]$, i.e. from 200 frames per sec to 8 frames per sec, 
and $a_{sig} \in [0.1,1]$ are performed and the fidelity
of the signals computed for optimal time-shifts. The results are shown
in the two color plots in Figure~\ref{fig:dyn} (C,D),
and some examples are shown in Panels (E) and (F).

Observe first from Panels (C,D) that
fidelity in both cases improves with signal strength and 
longer times between switches, with very good mean fidelity 
(of 5 degrees of less) for large $a_{sig}$ and $dt$. At the other end
of the spectrum,  there are hard lower bounds: fidelity is poor, to the point
that one could say the system fails, 
for signal strengths 
that are below a certain threshold or for refresh rates that are too fast.

Second, for the same values of $dt$ and $a_{sig}$, regular signals perform 
definitively better than random signals. Our model
is able to track with good accuracy rotating signals that turn very fast: $dt=10$
translates into 10 complete rotations per sec. At comparable frame rates, the
error is very large for random signals.

Turning to Panels (E,F), which show three examples from each of the two types
of signals, we see that when $dt$ is too small (e.g. 10 ms), the system is not only perpetually
``off" by some amount, it can miss certain features of the signal entirely. As $dt$ is increased, 
most of the errors captured by our fidelity metric 
are incurred through the rounding of corners; we think this is realistic,
and not necessarily undesirable. 
For large $dt$, as in the bottom panels in both (E) and (F),  
high fidelity means also that the read-out correctly reflects the piecewise constant 
nature of the signal, with visible overshoots depending on circumstance. 
In the case of the rotating signal, for example, the read-out shows
steps rather than a continuous motion when frame rate is too low. 
This seems realistic as well.

\bigskip
\noindent
{\bf Dynamical analysis}

\smallskip
Consider first the regularly rotating signal, as the situation there is more controlled:
 At $dt \in (10, 25)$, the signal is
essentially always in front of the peak of $R(\o,t)$ given that the time delay
discussed above is $20-30$ ms. Overshoots for this range of
$dt$ are irrelevant, and $R(\o,t)$ evolves in a fairly regular wave-like fashion, its front
chasing after and following closely the signal. The phenomenon
is consistent with that discovered in~\cite{BenYishai:1997p119} using
a rate model with a similar ring topology. 

For $dt$ too small, the wave can fall behind the signal,
as rise times of E-neurons do not permit infinitely fast movements of the front.
When the signal gets too far in front, the system starts to ``perceive" it as coming 
from behind. This explains the ``short-cuts" made by $\Oo(t)$ from one cycle 
to the  next by going backwards in the $dt=5$ plot in Panel (E) of Figure~\ref{fig:dyn} (see also~\cite{BenYishai:1997p119}).

For larger $dt$, say $dt > 60$ ms, the dynamical picture described 
in Section~\ref{overshoot} applies. One sees first an overshoot before the system corrects itself,
and pauses at the correct signal orientation for a noticeable duration before
continuing, as signal switches now occur significantly 
more slowly than the time it takes the wave
to go from one orientation to the next. Had we used larger jump sizes than
 the $\pi/10$ radian used here, the overshoots 
would have been even more pronounced. In any case, a step-like motion of  $\Oo(t)$
is expected.

\medskip 
Turning now to the random case, which models to some degree ``natural scenes",
a comparison of Panels (C) and (D) shows that in terms of fidelity it scores 
less well than the rotational case for nearly all values  of $a_{sig}$ and $dt$ 
and especially for small $dt$. We identify the following as contributing factors.

At least for larger $dt$, as in the bottom two plots in Panel (F), observe that a good fraction of the error is incurred at the larger jumps, which are absent in the rotational
case. At these jumps, averaging effects of our read-out function contributes to 
a lion's share of the rounding of corners, though network dynamics contribute
to that also (as can be deduced from the time constant in our read-out function).
The rounding of corners helps smooth out abrupt changes in scenery
(so it need not be an undesirable feature and is in fact likely to be realistic) but it does count against fidelity.

We have observed additional loss of fidelity when the signal turns around. Consider, for definiteness, the following scenario: Suppose we start with $\theta_{sig} = \theta_1$, and
the signal has stayed there long enough that spiking around $\theta_1$ is
solidly elevated. Then the signal switches to $\theta_2 > \theta_1$ and shortly thereafter
 to $\theta_3 < \theta_2$. 
As discussed in Section~\ref{switch_dyn}, the elevated I-spiking on
a wide interval around $\theta_1$  takes quite some time 
to resolve. If $\theta_3$ falls on this interval, its effect on the system can be nontrivially compromised. 

In more detail, if $dt$ is not too small, or if the signal remains near $\theta_3$ for 
long enough, the system will eventually overcome
the lingering suppression around $\theta_3$ and respond to the new signal. 
That is to say, it will track the signal, but with some additional delay (in addition
to that caused by the filter). A few examples of this phenomenon
 can be seen in the bottom two
plots of Panel (F), e.g. the faster tracking following
the signal switch at 0.3 sec and the slower one following the switch at 0.4 sec
in the bottom plot.

If $dt$ is too small, and the signal quickly moves away from $\theta_3$, then 
it can be missed altogether. Often, it is an untidy mix of delays and partially missed signals, many examples of which can be seen in the $dt=10$ ms 
plot in Panel (F). 

\medskip
To summarize, $R(\o,t)$ moves around in a wave-like fashion following the signal.
For large enough $dt$, the loss of fidelity is due in large measure to the filter, which
models heuristically what happens downstream. Focusing on our model network 
of V1, we observe that this early stage of processing already places limitations
on one's ability to track dynamic signals. Specifically, rise times for 
E-neurons impose an upper limit on the speed of the ``wave", equivalently 
a lower limit on $dt$, and this speed can be further
impeded by lingering suppression in its intended path when 
the signal makes a rapid turn-around. 

\bigskip \noindent
{\bf Comparison with frame rates in movies.}
Though we have tried to use biophysical constants in our modeling of V1,
we do not pretend that our filter is an accurate reflection of what goes on downstream,
and the precise values of fidelity depend on the filter. Thus our results should be
viewed as qualitative only. Nevertheless, the following comparison is interesting:
Television or movie projectors use frame rates of about 25 frames per sec~\cite{Efron:1973p1883,Thorpe:1996p1852}.
In our model, for strong random signals with $a_{sig}=1$ and $dt =40$, 
equivalently 25 frames per sec, fidelity is about 0.125, which means that 
the reconstructed signal has an error of about 0.125 radian (or 7-8 degrees) 
on average, these errors occurring mostly at the larger jumps in $\theta_{sig}$.
At 70-80 frames per sec, on the other hand, fidelity is poor, and
below about 10 frames per sec, one starts to perceive a visual signal as a series of still images~\cite{Boff:1986p2122,Thorpe:1996p1852}.
This is not inconsistent with our everyday experience, and suggests that 
perhaps our filter is not so far off.

\section*{Discussion}
The content of this paper can be summarized as follows: We constructed
a simple network of spiking neurons intended to model a single hypercolumn
of the visual cortex, and used it to study response properties to signals
in the form of time-varying orientations. To evaluate the network's
performance, we introduced two metrics: fidelity, which measures the accuracy
of the network's reconstructed signal, and reliability, which measures
trial-to-trial variability. While it is no surprise that there is a time lag
in the network's response following a signal switch, we found that this is
followed almost invariably by an overshoot, which can last over 100 ms
before the system corrects itself. Our analysis showed that this is a simple
consequence of phenomena that emerge as a result of 
the interaction between local E- and I-populations. 
We also compared regularly switching to randomly switching signals, and
found that the model can handle substantially faster switches in the first,
as can be predicted from the same dynamical analysis. As a final curiosity,
for random signals, frame rates that our model is able to track effectively
are comparable to those used in movies. These and other model predictions
can, in principle, be tested by measuring evoked responses in the real
visual cortex. 

{\it A conclusion of this work is that the same architecture of V1 that enables the
sharpening of orientation response through lateral interactions of E- and I-neurons 
is also responsible for limiting the speed at which orientation signals can be tracked 
and for producing certain artifacts in our perception.} This suggests that temporal limitations on visual perception occur very early on in the visual pathway.

A natural follow-up to this work is to study more general time-dependent
signals, such as moving stimuli. Indeed how the visual cortex computes
object velocities such as moving bars and moving patterns has been much
studied though far from understood (see e.g.~\cite{Lorenceau:1993p2113,Born:2005p2112,Smith:2005p2073,Movshon:1985p2070}).
Human perception (or misperception, as in illusions) of moving stimuli have
also been studied in multitudes of psychophysics experiments. The literature
offers few mechanistic explanation, however, for how the brain processes
visual information. For studies that involve higher cortical areas such as
MT and beyond, this is likely out of the reach at the present time. But much
of visual processing (of the local kind) is in fact carried out in V1 though not
completed there. We propose that this initial stage of processing can be
analyzed on the level of neuronal interactions. A model similar to ours,
made more realistic and enlarged to model a wider region of visual field,
can be used to study early-stage processing of moving signals that have
both orientation and spatial components, and metrics for quantifying
system performance, such as those introduced here, are clearly generalizable.
It is our hope that the present paper will inspire studies along these lines.
\section*{Appendix}

\subsection*{Numerical methods}
Simulations were implemented using Python and
Cython programming languages.
A standard Euler-Marayuma~\cite{Asmussen:2007p222} numerical integration scheme
was used to estimate solutions of
Equation~\eqref{sys}, treated as a stochastic
differential equation because of the term $I_{bkgd}^i$. Pseudo-random numbers used as noise increments were generated using the Mersenne Twister algorithm. A time-step of $\Delta t=0.05$ ms was used for all simulations. We verified that finer temporal resolution did not affect our results. 

Computation of statistical quantities such as filtered network responses, $Fid$ and $Rel$ were implemented in the MATLAB framework.

\subsection*{Model parameters}
\label{Params}
Parameters used to simulate System~\eqref{sys} are listed in the table below.
\small
\begin{center}
\begin{tabular}{|l|l|l|}
\hline
{\bf Param} & {\bf Value} & {\bf Description} \\
\hline\hline
$N$ & 1000 & total number of neurons \\
\hline
$C$ & 1 ($\mu$ F/cm$^2$) & membrane capacitance \\
\hline
$g_L$ & 0.1 (mS/cm$^2$)& leak conductance \\
\hline
$E_L$ & -65.0 (mV) & leak reversal potential \\
\hline
$\D T$ & 3.48 (mV) & spike slope factor \\
\hline
$V_T$ & -59.9 (mV) & threshold voltage \\
\hline
$T_{ref}$ & 1.7 (msec) & refractory period \\
\hline
$V_r$ & -68.0 (mV) & reset voltage \\
\hline
$\mu$ & 0.2 & background noise mean \\
\hline
$\e$ & 1.0 & background noise standard dev \\
\hline
\hline
$T^{EE}_{d}$ & 3.5 (msec) & spike delay for E $\to$ E synapses \\
\hline
$T^{IE}_{d}$ & 2.0 (msec) & spike delay for E $\to$ I synapses \\
\hline
$T^{EI}_{d}$ & 2.0 (msec) & spike delay for I $\to$ E synapses \\
\hline
$T^{II}_{d}$ & 2.0 (msec) & spike delay for I $\to$ I synapses \\
\hline
\hline
$a_{sig}$ & from 0 to 1.0 & external orientation signal center strength \\
\hline
$\s_{sig}$ & $\pi$/10 & external orientation signal width \\
\hline
\hline
$\s_{EE}$ & $\pi$/6 & E to E connectivity range standard dev \\
\hline
$\s_{IE}$ & $\pi$/6 & E to I connectivity range standard dev \\
\hline
$\s_{EI}$ & $\pi$/10 & I to E connectivity range standard dev \\
\hline
$\s_{II}$ & $\pi$/10 & I to I connectivity range standard dev \\
\hline
$p_{EE}$ & 0.15 & E to E connectivity probability (if in range) \\
\hline
$p_{IE}$ & 0.5 & E to I connectivity probability (if in range) \\
\hline
$p_{EI}$ & 0.5 & I to E connectivity probability (if in range) \\
\hline
$p_{II}$ & 0.5 & I to I connectivity probability (if in range) \\
\hline
\hline
$w_{EE}$ & 1.0  & E to E coupling \\
\hline
$w_{IE}$ & 0.85 & E to I coupling \\
\hline
$w_{EI}$ & -0.75 & I to E coupling \\
\hline
$w_{II}$ & -0.85 & I to I coupling \\
\hline
$\rho^{nmda}_E$ & 0.25 & NMDA current fraction of EE synapses  \\
\hline
$\rho^{nmda}_I$ & 0.5 & NMDA current fraction of IE synapses \\
\hline
$\tau_{EE}$ & 2.0 (msec) & E to E synaptic time-constant \\
\hline
$\tau_{IE}$ & 2.0 (msec) & E to I synaptic time-constant \\
\hline
$\tau_{EI}$ & 7.0 (msec) & I to E synaptic time-constant \\
\hline
$\tau_{II}$ & 7.0 (msec) & I to I synaptic time-constant \\
\hline
$\tau_{E-nmda}$ & 100.0 (msec) & NMDA to E synaptic time-constant \\
\hline
$\tau_{I-nmda}$ & 100.0 (msec) & NMDA to I synaptic time-constant \\
\hline
\end{tabular}
\end{center}


\begin{thebibliography}{10}

\bibitem{abbott1999neural}
L~Abbott and TJ~Sejnowski.
\newblock {\em Neural Codes and Distributed Representations: Foundations of
  Neural Computation}.
\newblock 1999.

\bibitem{Amarasingham:2006p1914}
A~Amarasingham.
\newblock Spike count reliability and the poisson hypothesis.
\newblock {\em Journal of Neuroscience}, 26(3):801--809, Jan 2006.

\bibitem{Asmussen:2007p222}
S~Asmussen and P~Glynn.
\newblock Stochastic simulation: Algorithms and analysis.
\newblock {\em Springer}, Jan 2007.

\bibitem{Beaulieu:1992p1763}
C~Beaulieu, Z~Kisvarday, P~Somogyi, M~Cynader, and A~Cowey.
\newblock Quantitative distribution of gaba-immunopositive and-immunonegative
  neurons and synapses in the monkey striate cortex (area 17).
\newblock {\em Cerebral Cortex}, 2:295--309, Dec 1992.

\bibitem{BenYishai:1995p1669}
R~Ben-Yishai, R~L Bar-Or, and H~Sompolinsky.
\newblock Theory of orientation tuning in visual cortex.
\newblock {\em Proc Natl Acad Sci USA}, 92(9):3844--8, Apr 1995.

\bibitem{BenYishai:1997p119}
Rani Ben-Yishai, David Hansel, and Haim Sompolinsky.
\newblock Traveling waves and the processing of weakly tuned inputs in a
  cortical network module.
\newblock {\em Journal of computational neuroscience}, 4(1):57--77, 1997.

\bibitem{Boff:1986p2122}
K~Boff, L~Kaufman, and J~Thomas.
\newblock Handbook of perception and human performance.
\newblock {\em Wiley-Interscience}, (ISBN 10: 0471885444 ISBN 13:
  9780471885443), Dec 1986.

\bibitem{Born:2005p2112}
R.~T Born, CC~Pack, CR~Ponce, and S~Yi.
\newblock Temporal evolution of 2-dimensional direction signals used to guide
  eye movements.
\newblock {\em Journal of Neurophysiology}, 95(1):284--300, Sep 2005.

\bibitem{Bressloff:2000p1671}
P~C Bressloff, N~W Bressloff, and J~D Cowan.
\newblock Dynamical mechanism for sharp orientation tuning in an
  integrate-and-fire model of a cortical hypercolumn.
\newblock {\em Neural Computation}, 12(11):2473--511, Oct 2000.

\bibitem{Burr:2011p2180}
David Burr and Peter Thompson.
\newblock Motion psychophysics: 1985-2010.
\newblock {\em Vision Res}, 51(13):1431--1456, Jun 2011.

\bibitem{Efron:1973p1883}
Robert Efron.
\newblock Conservation of temporal information by perceptual systems.
\newblock {\em Perception {\&} Psychophysics}, 14(3):518--530, Dec 1973.

\bibitem{Faisal:2008p1563}
A~Faisal, L~Selen, and D~Wolpert.
\newblock Noise in the nervous system.
\newblock {\em Nature Reviews Neuroscience}, Jan 2008.

\bibitem{Fitzpatrick:1985p1776}
D~Fitzpatrick, J~Lund, and G~Blasdel.
\newblock Intrinsic connections of macaque striate cortex: afferent and
  efferent connections of lamina 4c.
\newblock {\em J Neurosci}, Dec 1985.

\bibitem{FourcaudTrocme:2003p552}
Nicolas Fourcaud-Trocm{\'e}, David Hansel, Carl van Vreeswijk, and Nicolas
  Brunel.
\newblock How spike generation mechanisms determine the neuronal response to
  fluctuating inputs.
\newblock {\em Journal of Neuroscience}, 23(37):11628--40, Dec 2003.

\bibitem{Gabbiani:2009p47}
F~Gabbiani and C~Koch.
\newblock Principles of spike train analysis.
\newblock {\em Methods in Neural Modeling}, pages 313--360, Dec 2009.

\bibitem{Georgopoulos:1986p820}
A~Georgopoulos, A~Schwartz, and R~Kettner.
\newblock Neuronal population coding of movement direction.
\newblock {\em Science}, 233, Dec 1986.

\bibitem{Gray:1989p1811}
C~Gray, P~K{\"o}nig, A~Engel, and W~Singer.
\newblock Oscillatory responses in cat visual cortex exhibit inter-columnar
  synchronization which reflects global stimulus properties.
\newblock {\em Nature}, 338:334--337, Dec 1989.

\bibitem{Grunze:1996p2185}
H~C Grunze, D~G Rainnie, M~E Hasselmo, E~Barkai, E~F Hearn, R~W McCarley, and
  R~W Greene.
\newblock Nmda-dependent modulation of ca1 local circuit inhibition.
\newblock {\em J Neurosci}, 16(6):2034--43, Mar 1996.

\bibitem{Hawken:1996p2188}
M~J Hawken, R~M Shapley, and D~H Grosof.
\newblock Temporal-frequency selectivity in monkey visual cortex.
\newblock {\em Vis Neurosci}, 13(3):477--92, Jan 1996.

\bibitem{Henrie:2005p1830}
J~Henrie and R~Shapley.
\newblock Lfp power spectra in v1 cortex: the graded effect of stimulus
  contrast.
\newblock {\em Journal of Neurophysiology}, Jan 2005.

\bibitem{Holmgren:2003p2182}
C~Holmgren, T~Harkany, B~Svennenfors, and Y~Zilberter.
\newblock Pyramidal cell communication within local networks in layer 2/3 of
  rat neocortex.
\newblock {\em The Journal of Physiology}, 551(1):139--153, Aug 2003.

\bibitem{Hubel:1959p1933}
D~Hubel and T~Wiesel.
\newblock Receptive fields of single neurones in the cat's striate cortex.
\newblock {\em The Journal of physiology}, 148:574--591, 1959.

\bibitem{Hubel:1962p1940}
D~Hubel and T~Wiesel.
\newblock Receptive fields, binocular interaction and functional architecture
  in the cat's visual cortex.
\newblock {\em The Journal of physiology}, 160:106--154, 1962.

\bibitem{Hubel1995Eye}
David~H Hubel.
\newblock {\em Eye, Brain, and Vision (Scientific American Library, No 22)}.
\newblock 1995.

\bibitem{Kang:2003p1667}
Kukjin Kang, Michael Shelley, and Haim Sompolinsky.
\newblock Mexican hats and pinwheels in visual cortex.
\newblock {\em Proc Natl Acad Sci USA}, 100(5):2848--53, Mar 2003.

\bibitem{Koulakov:2002p2187}
Alexei~A Koulakov, Sridhar Raghavachari, Adam Kepecs, and John~E Lisman.
\newblock Model for a robust neural integrator.
\newblock {\em Nature Neuroscience}, 5(8):775--782, Aug 2002.

\bibitem{Laing:2001p1731}
C~R Laing and C~C Chow.
\newblock Stationary bumps in networks of spiking neurons.
\newblock {\em Neural Computation}, 13(7):1473--94, Jun 2001.

\bibitem{Lajoie:2013p785}
Guillaume Lajoie, Kevin~K Lin, and Eric Shea-Brown.
\newblock Chaos and reliability in balanced spiking networks with temporal
  drive.
\newblock {\em Phys. Rev. E}, 87(5):052901, May 2013.

\bibitem{Lin:2009p350}
Kevin~K Lin, Eric Shea-Brown, and Lai-Sang Young.
\newblock Spike-time reliability of layered neural oscillator networks.
\newblock {\em J Comput Neurosci}, 27(1):135--160, Aug 2009.

\bibitem{Lisman:1998p2184}
J~E Lisman, J~M Fellous, and X~J Wang.
\newblock A role for nmda-receptor channels in working memory.
\newblock {\em Nature Neuroscience}, 1(4):273--5, Jul 1998.

\bibitem{Lorenceau:1993p2113}
J~Lorenceau, M~Shiffrar, N~Wells, and E~Castet.
\newblock Different motion sensitive units are involved in recovering the
  direction of moving lines.
\newblock {\em Vision Res}, Dec 1993.

\bibitem{Movshon:1985p2070}
JA~Movshon, EH~Adelson, MS~Gizzi, and WT~Newsome.
\newblock The analysis of moving visual patterns.
\newblock {\em Pontificiae Academiae Scientiarum Scripta Varia}, 54:117--151,
  Apr 1985.

\bibitem{Oswald:2009p2183}
A.-M.~M Oswald, B~Doiron, J~Rinzel, and A.~D Reyes.
\newblock Spatial profile and differential recruitment of gabab modulate
  oscillatory activity in auditory cortex.
\newblock {\em Journal of Neuroscience}, 29(33):10321--10334, Aug 2009.

\bibitem{Pack2008189}
CC~Pack and RT~Born.
\newblock Cortical mechanisms for the integration of visual motion - (chapter
  2.11 in) the senses: A comprehensive reference.
\newblock pages 189 -- 218. 2008.

\bibitem{Priebe:2008p1675}
N~Priebe and D~Ferster.
\newblock Inhibition, spike threshold, and stimulus selectivity in primary
  visual cortex.
\newblock {\em Neuron}, Dec 2008.

\bibitem{Ringach:2002p1803}
D~Ringach.
\newblock Spatial structure and symmetry of simple-cell receptive fields in
  macaque primary visual cortex.
\newblock {\em Journal of neurophysiology}, Dec 2002.

\bibitem{Ringach:1997p1916}
D~Ringach, M~Hawken, and R~Shapley.
\newblock Dynamics of orientation tuning in macaque primary visual cortex.
\newblock {\em nature}, 387, 1997.

\bibitem{Ringach:2002p1805}
D~Ringach, M~Hawken, and R~Shapley.
\newblock Receptive field structure of neurons in monkey primary visual cortex
  revealed by stimulation with natural image sequences.
\newblock {\em Journal of vision}, Dec 2002.

\bibitem{Ringach:2002p1807}
D~Ringach, R~Shapley, and MJ~Hawken.
\newblock Orientation selectivity in macaque v1: diversity and laminar
  dependence.
\newblock {\em The Journal of Neuroscience}, 22(13):5639--5651, Dec 2002.

\bibitem{Ritt:2003p251}
Jason Ritt.
\newblock Evaluation of entrainment of a nonlinear neural oscillator to white
  noise.
\newblock {\em Phys. Rev. E}, 68(4):1--7, Oct 2003.

\bibitem{Rubin:2015p1673}
Daniel~B Rubin, Stephen D~Van Hooser, and Kenneth~D Miller.
\newblock The stabilized supralinear network: A unifying circuit motif
  underlying multi-input integration in sensory cortex.
\newblock {\em Neuron}, 85(2):402--417, 2015.

\bibitem{Salinas:1994p1840}
E~Salinas and L~Abbott.
\newblock Vector reconstruction from firing rates.
\newblock {\em J Comput Neurosci}, 1:89--107, Dec 1994.

\bibitem{Series:2004p575}
P~Seri{\`e}s, PE~Latham, and A~Pouget.
\newblock Tuning curve sharpening for orientation selectivity: coding
  efficiency and the impact of correlations.
\newblock {\em Nature neuroscience}, 7(10):1129--1135, 2004.

\bibitem{Shadlen:1998p481}
M~N Shadlen and W~T Newsome.
\newblock The variable discharge of cortical neurons: implications for
  connectivity, computation, and information coding.
\newblock {\em J. Neurosci.}, 18:3870--3896, 1998.

\bibitem{Smith:2005p2073}
Matthew~A Smith, Najib~J Majaj, and J~Anthony Movshon.
\newblock Dynamics of motion signaling by neurons in macaque area mt.
\newblock {\em Nature Neuroscience}, 8(2):220--228, Feb 2005.

\bibitem{Somers:1995p1672}
D~C Somers, S~B Nelson, and M~Sur.
\newblock An emergent model of orientation selectivity in cat visual cortical
  simple cells.
\newblock {\em J Neurosci}, 15(8):5448--65, Jul 1995.

\bibitem{Spiridon:2001p715}
Mona Spiridon and Wulfram Gerstner.
\newblock Effect of lateral connections on the accuracy of the population code
  for a network of spiking neurons.
\newblock {\em Network: Computation in Neural Systems}, 12(4):409--421, 2001.

\bibitem{Thorpe:1996p1852}
S~Thorpe, D~Fize, and C~Marlot.
\newblock Speed of processing in the human visual system.
\newblock {\em nature}, 381, Jun 1996.

\bibitem{Tiesinga:2008p462}
P~Tiesinga, JM~Fellous, and T~J Sejnowski.
\newblock Regulation of spike timing in visual cortical circuits.
\newblock {\em Nature Reviews Neuroscience}, 9(2):97--107, 2008.

\bibitem{vanVreeswijk:1996p476}
C~van Vreeswijk and H~Sompolinsky.
\newblock Chaos in neuronal networks with balanced excitatory and inhibitory
  activity.
\newblock {\em Science}, 274:1724--1726, 1996.

\end{thebibliography}


\end{document}